\documentclass[prl,showpacs,showkeys,twocolumn,10pt]{revtex4-1}
\usepackage{amsfonts,bbold,amsmath,amssymb,graphicx,epstopdf,verbatim,dsfont,color}
\usepackage[english]{babel}

\begin{document}

\title{The thermodynamics of time}
\author{Dries Sels}
\affiliation{TQC, Universiteit Antwerpen, Universiteitsplein 1, B-2610 Antwerpen, Belgium}
\author{Michiel Wouters}
\affiliation{TQC, Universiteit Antwerpen, Universiteitsplein 1, B-2610 Antwerpen, Belgium}
\date{\today}

\begin{abstract}

\end{abstract}
\maketitle
%\bibliographystyle{h-physrev}
%\bibliography{minfluct}

{\bf 
%Few things in physics are more elementary than time and yet we have little understanding of time as an entity itself. 
The problem of time is a deep paradox in our physical description of the world. 
According to Aristotle's relational theory, \emph{time is a measure of change} and does not exist on its own~\cite{aristotle}. In contrast, quantum mechanics, just like Newtonian mechanics, is equipped with a master clock that dictates the evolution of a system. This clock is infinitely precise and tacitly supplied free of charge from outside physics~\cite{wheeler1}. Not only does this absolute time make it notoriously difficult to make a consistent theory of quantum gravity, it is also the underlying problem in establishing the second law. Indeed, contrary to our experience, the Wheeler-deWitt equation~\cite{WdW,hartlehawking}  --a canonical quantization of general relativity-- predicts a static universe. Similarly, when simply concerned with the dynamics of a closed quantum system, there is no second law because the Von Neumann entropy is invariant under unitary transformations. Here we are mainly concerned with the latter problem and we show that it can be resolved by attributing a minimal amount of resources to the measurement of time. Although there is an absolute time in quantum mechanics, an observer can only establish a time by measuring a clock. For a \emph{local} measurement, the minimal entropy production is equal to the number of ticks. This lower bound is attained by a black hole.}

\begin{figure*}[tbp]
\centering
\includegraphics[width=\textwidth]{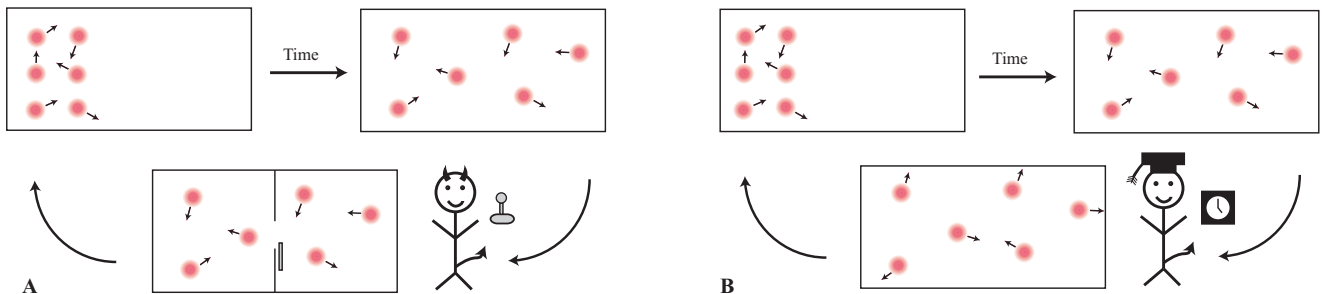}
\caption{ 
{\bf Quench dynamics. A} Maxwell's demon, who has access to the microscopic state of a classical gas, can put all particles on one side of a partition by timing the opening of a shutter such that the molecules preferentially go to one side. At the end of this process the demon has prepared a special state where all particles are on the left. By removing the partition the gas will expand under time evolution until it fills the whole container, after which the process can be repeated. {\bf B}  Loschmidt's demon can accomplish the same tasks without the need of the dividing wall. While the gas again expands in time from a state where all particles are on the left to a state where they fill the whole container, it is sufficient to invert all the velocities to make the system go back to its initial state. With knowledge of the initial condition and Hamiltonian of the system, the demon only requires a sufficiently accurate clock. Note that in order for the process {\bf A} to be cyclic, Maxwell's demon has to reset the record of his measurement whereas Loschmidt's demon in {\bf B} has to reset his clock.}
\label{fig:demon}
\end{figure*}

Reversible microscopic dynamics and a macroscopic arrow of time can be reconciled by taking into account the cost of information.
The connection between information and thermodynamics is established most directly in Maxwell's demon thought experiment~\cite{maxwell,maxrmp}, in which a demon with access to the microscopic state of a system can violate the second law, as illustrated in Fig.~\ref{fig:demon}. This, in turn, led Szilard to design an engine that could convert information into work~\cite{szilard}. Such a device was actually implemented by applying nonequilibrium feedback manipulation of a Brownian particle \cite{toyabe}.
The apparent paradox was not fully resolved until Landauer introduced the notion of logical irreversibility \cite{landauer}, by which he attributes a cost to the erasure of a memory. Maxwell's demon can therefore not beat the second law, as he has to erase the records of his measurements that were needed to open the shutter appropriately.

\begin{figure}[bp]
\centering
\includegraphics[width=0.45\textwidth]{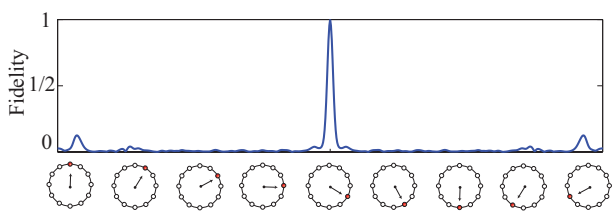}
\caption{{\bf Loschmidt echo.} Any operation that can reverse the state of the system has a fidelity that is sharply peaked around a certain point in time with a width of about the Boltzmann time $\tau_B$. In order to reverse the state of the system one must thus possess a clock with a resolution $\tau=\tau_b$. }
\label{fig:loschmidt}
\end{figure}

A similar analysis can be applied to the objection raised by Loschmidt to Boltzmann's proof~\cite{boltzmann} that the entropy always increases. Loschmidt argued that inverting all the velocities of the particles in a gas that had undergone some entropy increase, would bring the entropy back to its initial value, as illustrated in Fig.~\ref{fig:demon}.
It would be against the spirit of the second law to invoke any practical implementation aspects of the inversion of the velocities. Loschmidt could, for example, have fulfilled his task by giving each molecule a kick in the right direction. Again, it is required to know which kicks to apply. If we assume Loschmidt to know the initial condition of the gas, knowing the Hamiltonian is sufficient to know all positions and velocities at any time. We thus conclude that with a solution to the equations of motion and a sufficiently accurate {\em clock}, the inversion of the dynamics can be achieved, see Fig.~\ref{fig:loschmidt}.

Since the clock is a physical resource, it appears that the Loschmidt paradox would be solved if measuring time produces an entropy at least as large as the entropy reduction in the gas. Wigner already remarked that a clock is an essentially nonmicroscopic object \cite{wigner_rmp}, but we are not aware of a solution of the Loschmidt paradox in these terms.

A fundamental analysis of the entropy requires a quantum mechanical treatment. We therefore consider a quantum quench setting, where a system is prepared in a known quantum state $ | \psi_0 \rangle $, that is not an eigenstate of the Hamiltonian. Knowing the Hamiltonian allows to compute the state of the system any time 
\begin{equation}
| \psi(t) \rangle = e^{-i H t/\hbar} |\psi_0\rangle = \sum_n c_n e^{-i \epsilon_n t/\hbar}\, | n\rangle,
\end{equation}
where the states $| n\rangle$ are the eigenstates of the Hamiltonian with energy $\epsilon_n$ and $c_n=\langle n | \psi_0 \rangle $.

A definition of the entropy  for closed quantum systems, that has recently attracted a great amount of attention in relation to the second law is the so-called diagonal entropy~\cite{polkov_diag}. It is computed as the Shannon entropy of the diagonal elements of the density matrix in the instantaneous eigenbasis of the Hamiltonian. Here, it would read at all times after the quench $S_d = -\sum_n p_n \ln p_n$, with $p_n=|\langle n | \psi_0 \rangle |^2$.
It corresponds to the missing information for an observer that has no time information at all~\cite{SWGGE}, which is insufficient for our purposes. We thus have to define a proper time dependent entropy.

As a simple (but as we will show not to be physical) definition of the time dependent entropy, one could take the von Neumann entropy of the time averaged density matrix $S_d(t) = -{\rm Tr}[\bar \rho_t \ln ( \bar \rho_t)]$, where
\begin{equation}
\bar \rho_t = \frac{1}{t}\int_0^t \rho(t') dt'.
\end{equation}
In the limit for $t$ going to infinity, this entropy tends to the diagonal entropy. The initial increase of this entropy is logarithmic $S_d(t) \sim \ln (t/\tau_B)$, where the Boltzmann time scale $\tau_B$ depends on the energy width of the system as $\tau_B = \hbar/\Delta E$. It is the shortest time that occurs in the dynamics of the system and the minimal time required for the system to evolve to an orthogonal, distinguishable, state~\cite{ML}. 
For the Loschmidt scenario, this implies that a clock is required with a resolution $\tau=\tau_B$. The simplest wave function of such a clock is
\begin{equation}
|\mathcal C(t) \rangle = \sum_{k=1}^{n} e^{-i 2\pi k t/\tau } | k \rangle,
\label{eq:clock}
\end{equation}
It runs through $n$ pointer states over a time interval $t$ \cite{salecker}. Since the dimension of its microscopic Hilbert space is $n_{\rm ticks}=t/\tau$, the entropy produced by reading the clock is $\ln(t/\tau)$, which precisely coincides with the entropy decrease in the system. 
This reasoning illustrates how the Loschmidt paradox can be solved by incorporating the cost of timing.

While we believe this conclusion to be fundamentally correct, the logarithmic scaling of the entropy with time is not physical. Firstly, it does not allow a clock to be replaced by two clocks that run for half the time interval, which is in contradiction with a clock being a macroscopic object. Secondly, the logarithmic scaling of the entropy $S_d(t)$ does not correspond neither to the typical linear entropy production in nonequilibrium thermodynamics \cite{degroot}, nor to the linear increase of the entanglement entropy in quantum quenches \cite{eisert_rmp}.

The strongest phenomenological argument for a linear entropy production by a clock follows from the analysis by Salecker and Wigner \cite{salecker}. They have addressed the limits set by {\em locality} on the resources needed to construct a clock with arguments that are based on the Heisenberg uncertainty relation. They argue that for a readout of a clock after it has run over a time $t$ with accuracy $\tau$, it is necessary that the uncertainty in the position of the clock is less than $c \tau$, where $c$ is the speed of light. An initial spread in the position equal to $\lambda$, implies an uncertainty in its velocity of the order $\hbar / M \lambda$, when its total mass is $M$. The spread due to quantum diffusion $\hbar t / M \lambda$ limits the number of reliable ticks of the clock to \cite{salecker,ng}
\begin{equation}
n_{\rm ticks} = \frac{t}{\tau} \leq \frac{M c^2}{\hbar \tau^{-1}}.
\label{eq:nM}
\end{equation}
This relation shows that every tick of the clock requires an energy $\hbar \tau^{-1}$. Here, the mass clearly is a resource for the measurement of time. In view of time as information, this immediately implies a correspondence between `its' and `bits', between mass and information \cite{wheeler}.

The number of ticks of the clock can be directly related to entropy by using  general relativity. For a clock to tick at a time $\tau$, its physical dimension should not exceed $c \tau$.  If we want to avoid that our clock becomes a black hole and therefore unreadable, we need that $c \tau > r_S$ with the Schwarzschild radius $r_S=2 G M / c^2$. 
It is interesting to consider the limiting case where the clock becomes a black hole ($c \tau = r_S$), in which case \cite{ng,lloyd}
\begin{equation}
n_{\rm ticks}  \lesssim \frac{G M^2}{\hbar c} \sim S_{BH}.
\label{eq:nSBH}
\end{equation}
The Bekenstein-Hawking entropy $S_{BH}$~\cite{bekenstein,BCH,hawking} thus determines the number of ticks a black hole clock can make. In stark contrast to the abstract Hilbert space analysis which yielded a logarithmic entropy production, the relativistic phenomenology predicts an entropy production by the clock that scales linear with time, appropriate for a macroscopic classical object. Recall that the crucial additional ingredient in the phenomenological analysis is the concept of the locality.

The biggest problem with the standard quantum clock wave function \eqref{eq:clock} is that it does not include all the resources required for a realistic clock. It merely consists of a single harmonic oscillator degree of freedom. The only system that is fundamentally described by such a wave function is a freely propagating photon. The fact that one can never catch up with it renders it of course useless as a clock. In order to make a working clock based on photons, one needs some kind of a mirror, i.e. a massive object with many degrees of freedom.  A black hole can actually be seen as an implementation of a mirror, confining the photon on its horizon.

\begin{figure}[tbp]
\centering
\includegraphics[width=0.5\textwidth]{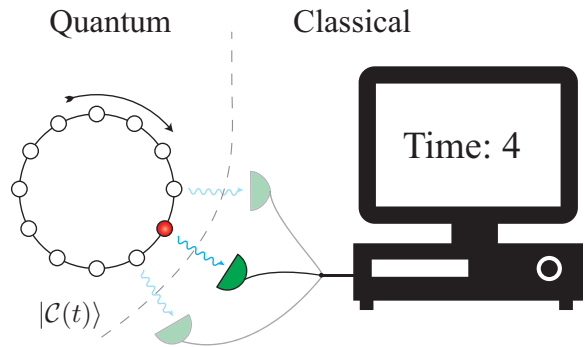}
\caption{ {\bf Clock.} A schematic representation of a clock. It consists of a particle hopping on a lattice. Each lattice site is coupled to a detector, that can see whether the particle has passed. The detectors can be read out by a classical computer which shows the time on a screen.}
\label{fig:clock}
\end{figure}

A complementary argument for the scaling \eqref{eq:nSBH}, consists of including in the clock wave function the minimal elements needed to measure the harmonic oscillator state, i.e. to make a clock that belongs to the macroscopic world. A schematic setup is sketched in Figure \eqref{fig:clock}. 
In order for the detectors to be able to reliably detect the time, their temperature  should satisfy $k_B T \leq \hbar \tau^{-1}$. Since, the clock needs an energy $E \geq n_{\rm ticks} \hbar \tau^{-1}$, we obtain that
\begin{equation}
n_{\rm ticks} \leq \frac{E}{T} = S.
\label{eq:nS}
\end{equation}
The black hole is merely a special case, where the temperature is given by the Hawking temperature $k_BT_H\sim \hbar c^3/GM$. The large value of the black hole entropy is the consequence of the fact that it evaporates by emitting a large amount of quanta at a very low energy, resulting in a very large number of clock ticks. 

It is actually remarkable that results \eqref{eq:nSBH} and \eqref{eq:nS} agree, even though they are based on a very different phenomenology. Where Eq. \eqref{eq:nSBH} was based on quantum uncertainty combined with general relativity, the derivation of Eq. \eqref{eq:nS} followed from an analysis of quantum measurement together with thermodynamics. We believe that this points towards a fundamental role for quantum measurement/information in the structure of space and time.

Because the Hawking temperature decreases with increasing mass, a faster clock is obtained for a lower mass. However, the lower the mass of the clock, the bigger its quantum diffusion. Reducing the mass too much leads to a clock that gets delocalised during a single period. The fastest clock that one can construct is a Planck mass black hole, which evaporates on a time of the order of the Planck time. 
Note the analogy between the evaporation of the Planck mass black hole on the Planck time scale and the decay of the Loschmidt echo. The signal shown in Fig. \ref{fig:loschmidt} can indeed be seen as a clock that ticks once. 

In conclusion we have shown that, similar to Maxwell's demon thought experiment, Loschmidt's paradox is resolved by taking into account the logical irreversibility. A demon that knows the laws of physics however only needs to know the time, which implies that knowledge of time comes at a cost. Including a  measurement apparatus, the number of ticks is limited by the entropy. When further constrained by general relativity we see that the number of bits  can not exceed the Bekenstein-Hawking entropy.

\acknowledgments
We thank J. Tempere, F. Brosens, J.T. Devreese and M. Richard for stimulating discussions. D.S. acknowledges support of the FWO as post-doctoral fellow of the Research Foundation - Flanders.

\end{document}